\documentclass[aps, prl, reprint,amsmath,amssymbols,superscriptaddress]{revtex4-1}

\usepackage{amsfonts}
\usepackage{graphicx}
\usepackage{bm}
\usepackage{color}
\usepackage{soul}
\usepackage{pdfpages}

\newcommand{\RR}{\mathbb{R}}

\begin{document}


\title{The weirdest martensite: Smectic liquid crystal microstructure and Weyl-Poincar{\'e} invariance}


\author{Danilo B. Liarte}
\email[]{dl778@cornell.edu}
\affiliation{Laboratory of Atomic and Solid State Physics, Cornell University, Ithaca, NY, USA}
\affiliation{Institute of Physics, University of S\~ao Paulo, S\~ao Paulo, SP, Brazil}
\author{Matthew Bierbaum}
\affiliation{Laboratory of Atomic and Solid State Physics, Cornell University, Ithaca, NY, USA}
\author{Ricardo A. Mosna}
\affiliation{Departamento de Matem{\'a}tica Aplicada, Universidade Estadual
de Campinas, 13083-859, Campinas, SP, Brazil}
\author{Randall D. Kamien}
\affiliation{Department of Physics and Astronomy, University of Pennsylvania,
Philadelphia, PA, USA}
\author{James P. Sethna}
\email[]{sethna@lassp.cornell.edu}
\affiliation{Laboratory of Atomic and Solid State Physics, Cornell University, Ithaca, NY, USA}


\date{\today}

\begin{abstract}

Smectic liquid crystals are remarkable, beautiful examples of materials microstructure, with ordered patterns of geometrically perfect ellipses and hyperbolas. The solution of the complex problem of filling three-dimensional space with domains of focal conics under constraining boundary conditions yields a set of strict rules, which are similar to the compatibility conditions in a martensitic crystal. Here we present the rules giving compatible conditions for the concentric circle domains found at two-dimensional smectic interfaces with planar boundary conditions. Using configurations generated by numerical simulations, we develop a clustering algorithm to decompose the planar boundaries into domains. The interfaces between different domains agree well with the smectic compatibility conditions. We also discuss generalizations of our approach to describe the full three-dimensional smectic domains, where the variant symmetry group is the Weyl-Poincar\'e group of Lorentz boosts, translations, rotations, and dilatations.

\end{abstract}


\maketitle

\begin{figure}[!ht]
\begin{minipage}[t]{0.69\linewidth}
\centering
\vspace{0.2cm}
(a) \par\smallskip
\includegraphics[width=0.8\linewidth]{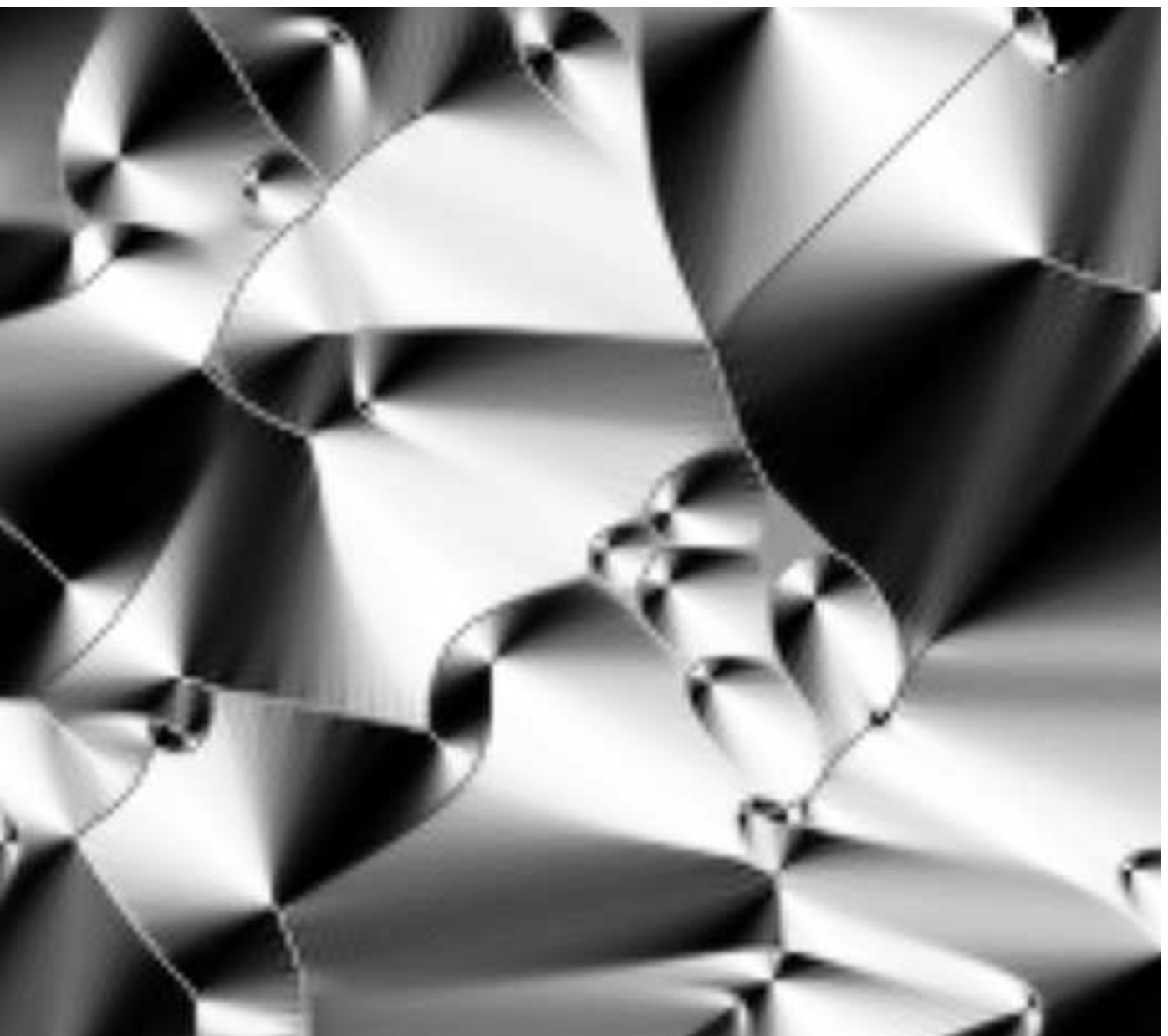}
\end{minipage}
\begin{minipage}[t]{0.29\linewidth}
\centering
(b) \par\smallskip
\includegraphics[width=\linewidth]{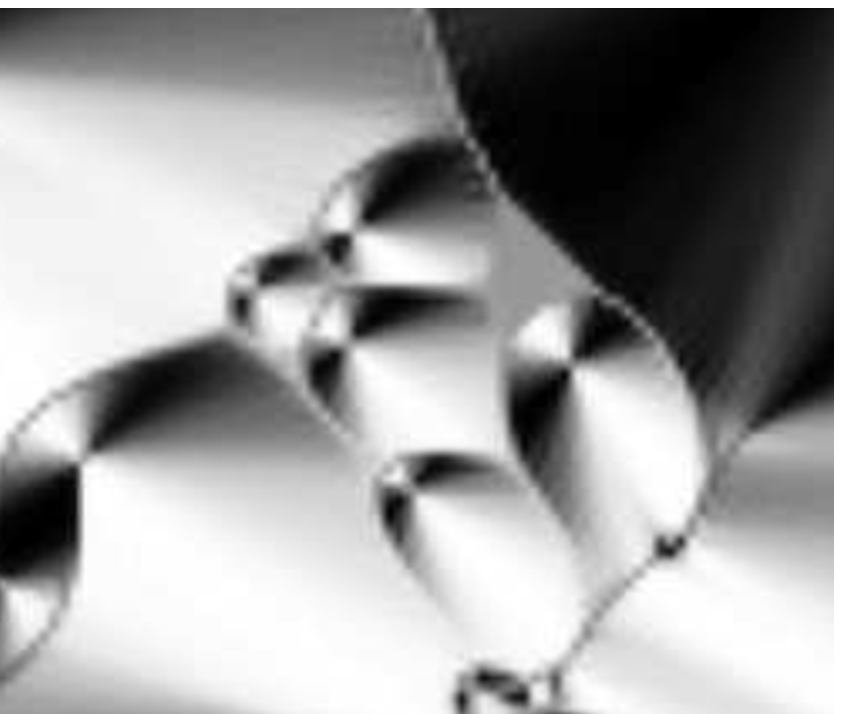}
\\
(c) \par\smallskip
\includegraphics[width=\linewidth]{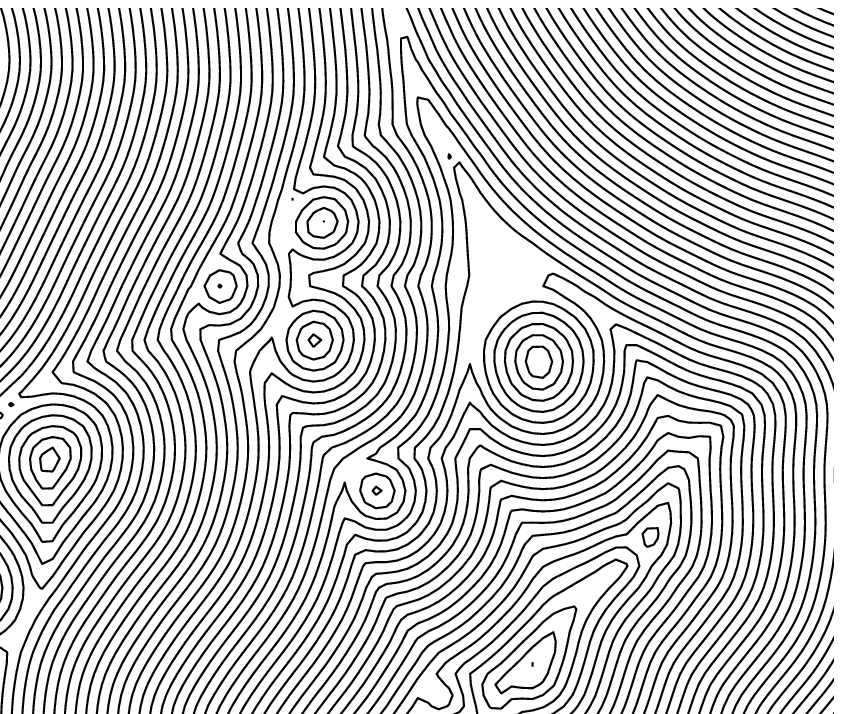}
\end{minipage}
\caption{(a) Crossed-polarizer images of focal conic domains on simulated smectics. (b) Elliptical defects attached to the upper surface (with planar boundary conditions). Note the hyperbolas emanating from the ellipses foci. Different focal conic domains (cyclides of Dupin) can rotate and deform via a Weyl-Poincar\'e transformation to join together compatibly continuous smectic layers. Note also the
small gaps between the ellipses. It is energetically favorable to fill these
regions with further ellipses, recursively down to molecular scales, leading
to the `Apollonian packing' microstructure.
(c) Layer sections forming concentric circles at the top boundary of the simulation.
\label{fig:polarizers}}
\end{figure}

The spatial decomposition of smectic liquid crystals into focal conic domains gives rise to one of the most unusual examples of materials microstructure. The smectic is a remarkable state of matter, breaking both the continuous rotational and (one-dimensional) translational symmetries of the isotropic fluid~\cite{gennes93}~\footnote{In this paper we ignore thermal fluctuations, which in three-dimensional smectic liquid crystals destroy the long-range translational order (leading to power-law correlations in the density-density correlation function perpendicular to the layers)~\cite{chaikin95}.}. In the beginning of the twentieth century, F. Grandjean and G. Friedel inferred that smectics were lamellar materials based on their bizarre microstructure~\cite{friedel10}; observe the beautiful patterns full of ellipses and hyperbolas in Fig.~\ref{fig:polarizers}. The figure shows a two-dimensional planar boundary (the layer surfaces lying perpendicular to the section) of a simulated configuration of a 3D smectic~A liquid crystal, mimicking experiments where thin slabs of smectic samples are placed between crossed polarizers~\footnote{Simulated crossed-polarizer images are produced via the calculation of the transmitted intensity of scattered light going through the anisotropic liquid-crystalline medium (see e.g. \cite{kleman03, gennes93}). To mimic focused images in the top slide of a smectic sample, we render a density plot with the gray scale corresponding to the scattering amplitude $N_x N_y$ at the top-layer section of a simulation. The layer surfaces in Fig~\ref{fig:polarizers}c) can be obtained by solving the equation $\bm{N}=\nabla \phi$ for $\phi$, using Fourier methods. More information on our smectic visualizers can be found in our previous publication~\cite{liarte15}.}. Friedel's breakthrough came with the realization that the visible conics could be modeled as the locus of the centers of curvature of a set of equally-spaced layers. The smectic layers must bend into cyclides of Dupin, which are the general set of equally-spaced surfaces whose singular centers of curvature lie along one-dimensional conics~\cite{hilbert99}. The smectic decomposes into the so-called focal conic domains (FCDs), which can be stabilized to mediate otherwise incompatible boundary conditions, such as anchoring of a sample boundary~\cite{kleman03}. 

We propose here a theory of smectic microstructure that generalizes and merges the laws of association between domains first proposed by G. Friedel~\cite{friedel22} and the mathematical theory of martensitic microstructure~\cite{bhattacharya03,ball87,ball92,ball04}. Our theory describes both the interpolation structure proposed by Beller et al.~\cite{beller13} to characterize the smectic flower textures, and Apollonius' packings in the FCD model of grain boundaries~\cite{kleman03}. Smectic liquids form the world's weirdest martensite.

In a martensitic transformation, the phase transition between different crystal structures (for instance from cubic to tetragonal symmetry) yields a low-temperature phase where two or more discrete configurations with different shape anisotropy coexist~\cite{bhattacharya03}. This structure  was discovered in c. 1890 by the microscopist Adolf Martens, though some of its mechanical properties have been used since (at least) 
the dawn of the Iron age. Metallurgists and blacksmiths manipulate the martensitic microstructure (as 
well as the dislocation and precipitate structures) by heating and hammering
swords and horseshoes to confer toughness and strength.

\begin{figure}[!ht]
\includegraphics[width=0.8\linewidth]{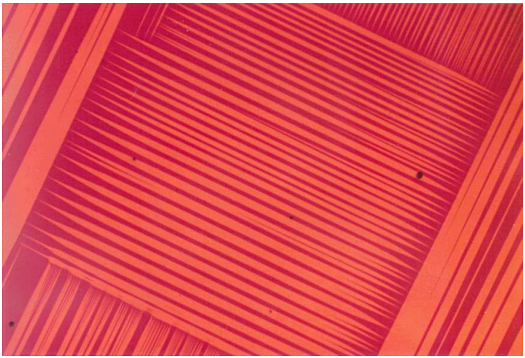}
\caption{Experimental martensitic microstructure. Different variants (colors) $U_i$ can rotate via an $\text{SO}(3)$ symmetry so as to joint together compatibly at twin boundaries. Courtesy of C.~Chu and R.~D.~James~\cite{chu93}.
\label{fig:martensite}}
\end{figure}

Martensites are usually characterized by a striped pattern, or laminate, that minimizes the constrained elastic free energy while keeping the net strain near zero. Fig.~\ref{fig:martensite} shows an example of a martensitic structure, with the dark and light regions representing two variants of the crystal martensite (see also SM discussion of paper folding as a martensite). The martensitic variants are akin to the smectic domains filled with a single family of Dupin cyclides.

In this paper, we generalize the mathematical theory of martensites in order to study the microstructure of smectic liquid crystals. We shall start
by labeling the energy-minimizing states and variant symmetry groups of smectic and martensitic crystals. We then extract smectic configurations from planar boundaries of our simulations and apply a clustering algorithm to decompose two-dimensional space into domains where the layers form sets of concentric circles (Fig.~\ref{fig:polarizers}c). We finish by a discussion of some physical examples and open questions.

In a martensitic phase of a cubic to tetragonal transformation, we can describe the system by the vector field $\bm{y}=\bm{y}(\bm{x})$, where $\bm{x}$ and $\bm{y}$ are the positions of a point in reference and target spaces, respectively, and the reference space is associated with the austenite configuration. The martensite variants are described by the gradient tensor $\nabla \bm{y} = (( \partial y_i / \partial x_j ))$, which can assume one of the three forms:
\begin{eqnarray}
& U_1 = \left( 
\begin{array}{ccc}
\eta^2 & 0 & 0 \\
0 & 1 / \eta & 0 \\
0 & 0 & 1 / \eta
\end{array}
\right), \quad 
U_2 = \left( 
\begin{array}{ccc}
1 / \eta & 0 & 0 \\
0 & \eta^2 & 0 \\
0 & 0 & 1 / \eta
\end{array}
\right),
\nonumber \\
&
U_3 = \left( 
\begin{array}{ccc}
1 / \eta & 0 & 0 \\
0 & 1 / \eta & 0 \\
0 & 0 & \eta^2
\end{array}
\right),
\label{eq:MartensiteVariants}
\end{eqnarray}
for a uniaxial volume-conserving stretch along the three cartesian axes. The set of energy-minimizing states consists of all possible rotations of the three deformation variants, and can be written as:
\begin{eqnarray}
K = \bigcup_{i=1}^3 \text{SO} (3) \cdot U_i,
\label{eq:MartensiteGroundStates}
\end{eqnarray}
where $\text{SO}(3)$ denotes the group of three-dimensional proper orthogonal transformations (rotations).

Similarly,
smectics can be described by a scalar displacement field $\phi=\phi(\bm{x})$, which measures the local displacements from a set of flat equally-spaced surfaces. The smectic layers are equipotential surfaces of $\phi$, with the layer
normals $\bm{N}=\bm{N}(\bm{x}) \equiv - \nabla \phi$.
Note that the displacement field defines a surface with constant slope ($|\nabla \phi|=1$) in the four-dimensional `space-time' ($\{\phi,x,y,z\}$) -- forming
`light surfaces' in the order parameter field (see Ref.~\cite{alexander10}). This
analogy to special relativity, and the Lorentz invariance of the allowed
smectic domains, will be central to our proposed martensitic analysis of 3D
smectic layers.

The Dupin cyclides may be defined as the surfaces whose centers of curvatures
lie along one-dimensional curves. Since the condition of equally spaced layers
($N^2 \equiv 1$) implies that the centers of curvature are shared by subsequent
surfaces, a domain filled with Dupin cyclides allows the system to form
relatively cheap line singularities, rather than the energetically expensive
two-dimensional singular centers of curvature of typical curved surfaces. 
The geometry of the Dupin cyclides furthermore forces the singular curves
to be conic sections -- generically ellipses and hyperbolas passing
perpendicularly through one another's foci. 

\begin{figure}[!ht]
\centering
\includegraphics[width=0.8\linewidth]{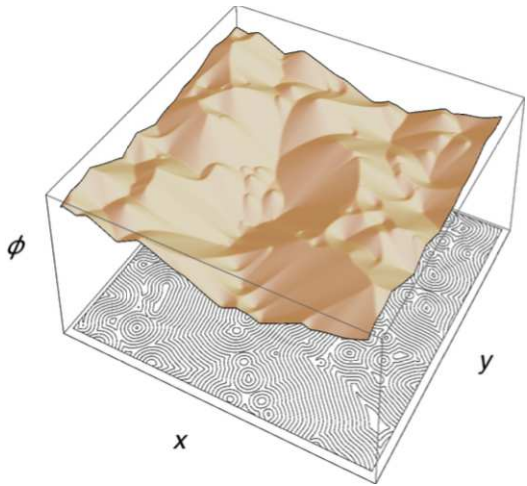}
\caption{
Displacement field $\phi$ (top grey surface) and some of its levels (bottom black lines) as a function of $x$ and $y$ for the top planar boundary of a simulated configuration of smectic liquid crystals. Note that each domain of concentric
circles on the bottom becomes a positive or negative light cone in the
surface defining $\phi(x,y)$.
\label{fig:phiPlots}}
\end{figure}

Before analyzing 3D smectic domains, let us analyze a simpler case: 
smectic layers at a flat interface,
where the layers are constrained to approach perpendicular to the boundary.
Such {\em planar} boundary conditions are often found at surfaces like glass
slides; they are called planar because the smectic molecules
(normal to the layers) are in the plane of the boundary surface.
Fig.~\ref{fig:phiPlots} shows the displacement field $\phi$ (top grey surface)
and some layers (bottom black lines) as a function of $x$ and $y$ at the
top (planar) boundary of a smectic configuration (the same used in
Fig.~\ref{fig:polarizers}). 
At a planar boundary, the cyclides of Dupin form concentric circles
(Fig.~\ref{fig:polarizers}c), corresponding to `light cones' in the figure
with space-time centers $\{\phi_0, x_0, y_0\}$.
The corresponding displacement fields can
be described in terms of two variants
\begin{equation}
U^{(2)}_\pm: \phi = \pm \sqrt{x^2 + y^2}
\label{eq:2dVariants}
\end{equation}
together with the three-dimensional group of translation operations (two translations in space and one in time)
\begin{equation}
T(3): \{\phi, x, y\} \to \{\phi-\phi_0, x-x_0, y-y_0\}
\end{equation}
leading to a space of low energy structures 
\begin{equation}
K^{(2)} = \bigcup_{\alpha=\pm} \text{T} (3) \cdot U^{(2)}_\alpha.
\label{eq:2dLowEnergy}
\end{equation}

In the full three-dimensional smectic domains, we may not only translate
and rotate the Dupin cyclide domains, but we may also transform them under
dilatations and Lorentz boosts (which change the eccentricity of the ellipses 
and hyperbolas~\cite{alexander10}, leading~\cite{alexander12} to 
the Weyl-Poincar\'e group $\mathcal{WP}$~\cite{bertrand90}. This group, which is a semi-direct product of positive dilatations and Poincar\'e transformations, is an 11 dimensional group. We can form a general Dupin domain by the action of 9 generators of $\mathcal{WP}$ (which correspond to the quotient of $\mathcal{WP}$ by a 2D Abelian subgroup) on a toroidal domain, whose singular curves are a unit circle in the $xy$ plane and a perpendicular line through its center (see section II in Supplemental Material). Since this domain is the product of two cones, $[ (r+1)^2 + z^2-\phi^2][(r-1)^2 + z^2-\phi^2]=0$, with $r=\sqrt{x^2+y^2}$, there are four variants now~\footnote{The two extra choices for the sign correspond to setting $(r-1)^2 + z^2-\phi^2=0$ or $(r+1)^2 + z^2-\phi^2=0$ in the equation for the toroidal FCD (see main text). In the first case we obtain Dupin surfaces formed by regions with both positive and negative Gaussian curvature; the FCD of the first species corresponds to the negatively curved regions. In the second case we obtain the FCD of the second species, for which only the hyperbola is visible under the microscope (the ellipse being virtual)~\cite{kleman03}.},
\begin{equation}
U^{(3)}_{\pm\pm}: \phi = \pm \sqrt{\left(r \pm 1\right)^2 + z^2},
\end{equation}
hence leading to a huge space of low energy structures 
\begin{equation}
K^{(3)} = \bigcup_{\alpha,\beta=\pm} \mathcal{WP} \cdot U^{(3)}_{\alpha\beta}.
\end{equation}

\begin{figure}[!ht]
\includegraphics[width=0.8\linewidth]{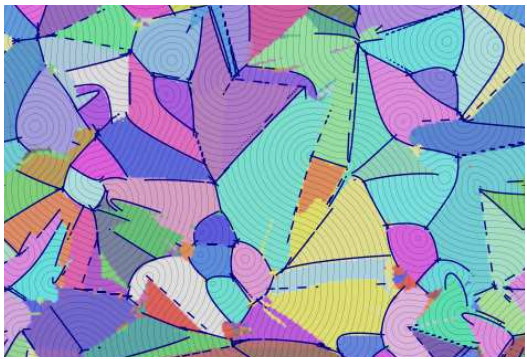}
\caption{Smectic microstructure of a two-dimensional planar boundary of a 3D simulation. The intersection of smectic layers and the section form sets of concentric circles, which are shown as thin black curves. The boundary between clusters (blue lines) are conics satisfying suitable compatibility conditions.
Straight dashed lines separate domains whose `light cones' are pointed
in opposite directions and just miss tangency.
\label{fig:SmecticMicrostructure}}
\end{figure}

We employ numerical simulations to generate the smectic configurations that are used in our microstructure analysis. Our simulations describe the dynamical evolution of the layer normal field $\bm{N}=\bm{N}(\bm{r})$ along the gradient-descent path of an elastic free energy~\cite{liarte15}. We consider the following adaptation of the Oseen-Frank free-energy functional~\cite{oseen33, zocher33, frank58}:
\begin{eqnarray}
F_{\text{s}} = \int d\bm{r} \left[ f_{\text{s}} (\bm{N}, \nabla \bm{N} ) + \bm{\lambda} \cdot \nabla \times  \bm{N} \right] ,
\label{eq:total_energy}
\end{eqnarray}
with the energy density $f_{\text{s}}$ given by
\begin{eqnarray}
f_{\text{s}} &=& \frac{ B }{4} (1- N^4)^2 + K N^2 \left( \nabla \cdot \bm{N} \right)^2 \nonumber \\
&& \quad + \frac{1}{2} K_{24} N^2 \nabla \cdot \left[\left(\bm{N} \cdot \nabla \right) \bm{N} - \bm{N} \left(\nabla \cdot \bm{N}\right)\right],
\label{eq:energy_density}
\end{eqnarray}
where $B$, $K$, and $K_{24}$ are constants. The first term in Eq.~\eqref{eq:energy_density} is a compression term, which penalizes elastic distortions of the smectic interlayer spacing. The second and third terms are related to the usual splay and saddle-splay distortions~\cite{gennes93}. Notice the unusual amplitude dependence ($\sim N^2$) multiplying the $K$ and $K_{24}$ elastic terms. It originates in gradient distortions of the form $(\nabla Q)^2$, which are proportional to $N^2$ for nematic uniaxial ordering~\cite{wright89, liarte15}, where $Q=((Q_{i\, j}))$ is the Maier-Saupe tensorial order parameter. We also use a Lagrange multiplier $\bm{\lambda}$ to forbid the existence of dislocations. The layer-normal field $\bm{N}$ satisfies the set of partial differential equations:
\begin{eqnarray}
\gamma \, \dot{ \bm{N} } = - \left( \frac{\delta F_{\text{s}}}{ \delta \bm{N}} - \left< \frac{\delta F_{\text{s}}}{ \delta \bm{N}} \right>  \right),
\label{dynamics}
\end{eqnarray}
where the angle brackets denote a spatial average and $\gamma$ is a viscosity constant. The second term of Eq.~\eqref{dynamics} ensures that the net number of layers in the cell does not change during a gradient descent step. Initially, we generate a random order parameter field $\bm{N}$, and use an Euler integrator with adaptive control system to solve the set of PDEs given by Eq.~\eqref{dynamics}. We consider a cubic grid with global tetragonal shape of size 256$\times$256$\times$64. The elastic constants are fixed so that de Gennes' length scale $\xi \equiv \sqrt{K/B} = 0.2 a$, and $K / K_{24} = -1.5$, where $a$ is the simulation lattice spacing. In the present paper, we only consider planar anchoring with the top and bottom boundaries, i.e. we fix $N_z=0$ at $z=0$ and $z=L_z$. Our code combines the versatility of Python with fast parallel programming using CUDA. To obtain the configuration displayed in Figs. \ref{fig:polarizers}, \ref{fig:phiPlots} and \ref{fig:SmecticMicrostructure}, we evolved $\bm{N}$ for a total time $t_{\text{t}}\approx 2,000 \, \gamma / B$~\footnote{In our simulations, the coarsening of FCDs does not stop until a flat configuration is reached (though it becomes very slow at later times). We have also applied our methodology to later configurations, which display larger and simpler features, but no essential qualitative changes. As discussed in~\cite{liarte15}, we are able to stabilize defect structures by simulating ``dust'' particles on the boundaries.}. More details of the simulations can be found in~\cite{liarte15}.

We developed a clustering algorithm to decompose smectic planar sections into domains with distinct centers, i.e. the low-energy structures described in Eqs. (\ref{eq:2dVariants}-\ref{eq:2dLowEnergy}).  For each domain (we start with square clusters), we use a least-squares optimization algorithm to find the four-tuple $X_{\alpha}=(\phi_{0, \alpha}, x_{0,\alpha}, y_{0,\alpha}, \sigma_{0,\alpha})$ that minimizes the cost function (see Supplemental Material)
\begin{eqnarray}
C_{\alpha} = \sum_{i } c \, ( r_{i,\alpha} , X_{\alpha}),
\end{eqnarray}
where the sum runs over points in ($2+1$)D space that belong to cluster $\alpha$, and
\begin{eqnarray}
&& c\, ( r_i , X_{\alpha})
 = \left\{
\phi_i - \phi_{0,\alpha}
+ \sigma_{0,\alpha}
\left[ \left(x_i - x_{0,\alpha} \right)^2 
\right. \right. \nonumber \\ && \quad \left. \left.
+ \left(y_i - y_{0,\alpha} \right)^2 \right]^{1/2} \right\}^2
+ \left( N_x - \frac{x_i - x_{0,\alpha}}{d_{i 0,\alpha}}\right)^2
\nonumber \\ && \quad
+ \left( N_y - \frac{y_i - y_{0,\alpha}}{d_{i 0, \alpha}}\right)^2,
\label{eq:Cost}
\end{eqnarray}
where $d_{i 0,\alpha}=[(x_i-x_{0,\alpha})^2+(y_i-y_{0,\alpha})^2]^{1/2}$ is the $2D$ Euclidean distance from point $i$ to the center, and $\sigma_{0,\alpha} =\pm 1$ characterizes the nappe of the cone~\footnote{The ``nappe'' of a double cone is the conic region above \emph{or} below the cone apex.}. By minimizing the first term in the right-hand side of Eq. \eqref{eq:Cost}, we find the best approximation for the local energy minimizer in the set \eqref{eq:2dLowEnergy}. The second and third terms make the analysis sensitive to gradient changes. The next step is to redefine the clusters so that each pixel in ($2+1$)D space is associated with the center that yields the least cost $c(r_i,X_\alpha)$. This entire process is iterated several times. At the end, we merge a few clusters that are described by similar parameters $X_\alpha$. Fig.~\ref{fig:SmecticMicrostructure} shows a plot of our cluster decomposition, where each pixel is colored according to the cluster centers $X_\alpha$.

The colored regions in Fig.~\ref{fig:SmecticMicrostructure} are analogous to the martensitic domains in Fig.~\ref{fig:martensite}. Note the elliptical domains (just lower left of center) are only a small fraction of the total area; as in many experiments, much of the region is not an Apollonian packing of focal conic domains (and hence not described by Friedel's laws of association). The non-elliptical domains nonetheless appear to be filled with cyclides of Dupin.

Finally, we turn to the compatibility conditions. In our example of
martensites (Eqs.~\ref{eq:MartensiteGroundStates}
and~\ref{eq:MartensiteVariants}), the constraint that the field
$\mathbf{y}(\mathbf{x})$ is continuous forces the boundaries between 
variants to be rotated by specific angles: if
$\nabla \mathbf{y} = K_1 = R_1 U_i$ and
$K_2 = R_2 U_j$ are to meet continuously along a twin boundary (the boundaries
between the lamellae in Fig.~\ref{fig:martensite}), then
$K_1 - K_2$ must be zero along the boundary.

What are the compatibility conditions for our smectic concentric sphere 
domains at the planar boundary, allowing $\phi(\mathbf{y})$ to be continuous? Compatible boundaries between domains of concentric circles are the projections of the intersection of two conic surfaces onto the $xy$-plane. We can find the singular conic solutions by solving the pair of equations:
\begin{eqnarray}
&& (x-x_0)^2+(y-y_0)^2 - (\phi-\phi_0)^2 = (x-x_1)^2
\nonumber \\ 
&& \quad +(y-y_1)^2 - (\phi-\phi_1)^2,
\end{eqnarray}
\begin{eqnarray}
(x-x_0)^2+(y-y_0)^2 - (\phi-\phi_0)^2 = 0,
\end{eqnarray}
for $x$ and $y$. This algebraic manipulation results in the smectic compatibility equation:
\begin{eqnarray}
&& \left( x - x_0 \right)^2 + \left( y - y_0 \right)^2 = \left[x_0^2 - x_1^2 - 2x(x_0 - x_1) + y_0^2
\right. \nonumber \\ &&  \left. \quad
 -y_1^2 - 2y(y_0 - y_1)
+ (\phi_0 - \phi_1)^2\right]^2 / \left[4 (\phi_0 - \phi_1)^2\right],
\label{eq:compatibility}
\end{eqnarray}
a quadratic equation whose solutions are conic sections in the boundary
surface dividing neighboring smectic domains.
The dark lines in Fig.~\ref{fig:SmecticMicrostructure} show compatible boundaries in the smectic microstructure given by the ellipses and hyperbolas of Eq.~\eqref{eq:compatibility}. 

We conjecture that the compatibility condition for 3D smectic domains will
lead to boundaries, as described by Friedel~\cite{friedel22}, which are portions
of right circular cones connecting one point of a conic to its confocal partner.

How have we altered the standard theory of martensites?
Firstly, our elastic free energy density is written in terms of the
gradients of $N$, and hence {\em second} derivatives of the displacement
field $\phi$ -- a strain gradient theory. Second, our domains are not
described by a uniform deformation, but rather by a deformation determined
by the non-local constraints~\cite{sethna82b} imposed by the constraint
of one-dimensional singularies. Conversely, what further can we glean from the martensitic analogy?
The mathematical engineers use sophisticated real analysis (minimizing 
sequences and Young measure distributions) to describe the family of
boundary conditions that can be relaxed by an infinitely fine
microstructure~\cite{ball87,ball92,ball04}. Apollonian microstructures
formed by a hierarchy of ellipses are known to mediate smectic tilt
boundaries~\cite{kleman03,kleman00,bidaux73}.
An infinitely fine laminate of alternating concentric spheres and
Dupin domains, inspired by the experimental `flower texture', has been shown
to relax an arbitrary cylindrically symmetric boundary
condition~\cite{beller13}. But the general question remains a fascinating
one: what is the class of smectic boundary conditions that can be mediated
by structures of equally spaced layers with only line singularities? 

\begin{acknowledgments}
We would like to thank R. D. James for useful conversations. DBL acknowledges the financial support provided by the Brazilian agency Capes. This work was supported in part by the Department of Energy DOE-BES DE-FG02-07ER46393 (JPS, MKB, and DBL) and by a Simons Investigator grant from the Simons Foundation to RDK. RAM and RDK were supported through NSF Grant DMR12-62047. R.A.M. acknowledges financial support from FAPESP grant 2013/09357-9.
\end{acknowledgments}

%

\pagebreak
\widetext
\begin{center}
\textbf{\large Supplemental Material to: \\ The weirdest martensite: Smectic liquid crystal microstructure and Weyl-Poincar{\'e} invariance}
\end{center}
\setcounter{equation}{0}
\setcounter{figure}{0}
\setcounter{table}{0}
\setcounter{page}{1}
\makeatletter
\renewcommand{\theequation}{S\arabic{equation}}
\renewcommand{\thefigure}{S\arabic{figure}}
\renewcommand{\bibnumfmt}[1]{[S#1]}
\renewcommand{\citenumfont}[1]{S#1}

In this text, we present a brief introduction to the mathematical theory of martensites. We also expand the argument for the four-dimensional low-energy structures of smectics, and give additional details on our numerical domain decomposition.

\section{An introduction to martensites and paper folding}

Some crystals undergo a solid-to-solid phase transformation upon decreasing temperature, the \emph{martensitic transformation}, characterized by an abrupt change in the crystalline lattice structure~\cite{bhattacharya03}. Above the transition temperature, the crystal is in the austenite phase, with a lattice unit cell that typically has greater crystallographic symmetry (such as cubic). In the low-temperature, \emph{martensite phase}, two or more symmetry-related \emph{variants} coexist; for instance, in a cubic-to-tetragonal transformation, the low-temperature variants correspond to tetragonal unit cells with the symmetry axes parallel to each of the three cartesian axes. The transformation is both reversible and \emph{self-accommodating}, i.e. it does not involve a net change in the material shape. A fascinating microstructure originates in the self-accommodating property: martensite variants form fine laminates, compatibly merged together along twin boundaries. The martensite microstructure, along with the fact that there is only one variant of austenite, ultimately lead to very interesting behavior, such as the shape-memory effect~\cite{bhattacharya03}. 

A beautiful theory of the martensitic microstructure is based on a nonlinear elasticity model~\cite{ball87, ball92, ball04}. This theory combines techniques extracted from several branches of mathematics, from nonlinear analysis and the calculus of variations to partial differential equations and geometry. The experimental existence of crystal laminates, where the deformation gradient jumps when crossing a boundary across a smooth surface, makes room for the use of functions where the partial derivatives are only required to exist in a weak sense. It is then convenient to use functions that are elements of a Sobolev space~\cite{evans10}, $W^{1,1}=W^{1,1}(\Omega, \mathbb{R}^3)$, where $\Omega$ is a bounded region of $\mathbb{R}^3$. The order parameter is the vector field $y(x): \Omega \rightarrow \mathbb{R}^3$, giving the new (three-dimensional) position after a deformation, i.e. the position in target space, and such that
\begin{eqnarray}
\int_\Omega \left[ \left| y(x) \right| + \left| \nabla y (x) \right| \right] dx < \infty,
\end{eqnarray}
where the notation $\nabla y$ represents the deformation gradient matrix $\nabla y = ((\partial y_i / \partial x_j ))$.

The mathematical problem consists in finding the minimum of the total free energy functional of the crystal:
\begin{eqnarray}
F_{\text{m}} \left[y(x)\right] = \int_\Omega f_{\text{m}} \left( \nabla y\right) d x,
\end{eqnarray}
where the free energy density $f_{\text{m}}$ also depends on the temperature $T$, and suitable boundary conditions are imposed. Below the transition temperature, the set of energy-minimizing deformation gradients $K=K(T)$ can be written as:
\begin{eqnarray}
K(T<T_c) = \bigcup_{i=1}^N \text{SO} (3) \, U_i (T),
\end{eqnarray}
where $U_i (T)$ are the deformation gradients of the $N$ variants of martensite. For instance, $N=3$ for a cubic-to-tetragonal transformation, with the variant strains characterizing volume-conserving uniaxial deformations along each cartesian direction:
\begin{eqnarray}
& U_1 = \left( 
\begin{array}{ccc}
\eta^2 & 0 & 0 \\
0 & 1 / \eta & 0 \\
0 & 0 & 1 / \eta
\end{array}
\right), \quad 
U_2 = \left( 
\begin{array}{ccc}
1 / \eta & 0 & 0 \\
0 & \eta^2 & 0 \\
0 & 0 & 1 / \eta
\end{array}
\right), \quad
&
U_3 = \left( 
\begin{array}{ccc}
1 / \eta & 0 & 0 \\
0 & 1 / \eta & 0 \\
0 & 0 & \eta^2
\end{array}
\right).
\label{eq:MartensiteVariants}
\end{eqnarray}

Now consider a surface $S$ separating two crystal variants, so that $\nabla y$ is continuous on either side of $S$, and the limits $\nabla^+ \, y (z) = A$ and $\nabla^- \, y (z) = B$ for a point $z \in S$, and $A, B \in K(T<T_c)$. To ensure that the tangential derivatives are equal at $z$, and that the deformation field is continuous at the boundary, one uses the compatibility condition (or Hadamard jump condition):
\begin{eqnarray}
A-B = \bm{a} \bm{n},
\label{eq:hadamard_condition}
\end{eqnarray}
where $\bm{n}$ is a vector that is normal to the surface $S$, $\bm{a} \in \mathbb{R}^3$, and $\bm{a} \bm{n} = (( a_i n_j))$ is a $3\times 3$ matrix.

The lack of convexity (strictly speaking: \emph{quasiconvexity}) of $f_{\text{m}} ( \nabla y )$, suggests that the minimum of the free energy functional $F_{\text{m}} [ y(x) ]$ is not attained. One can assume without loss of generality that $f_{\text{m}}(A)=0$, for $A \in K$, and that the infimum of the total free energy is also zero. Then it is possible to construct a minimizing sequence $y^{(j)}$ for $F_{\text{m}}$ for which $\lim_{j\rightarrow \infty} F_{\text{m}} [ y^{(j)} ] = 0$, and such that the Young measure $\nu_x$, $x \in \Omega$, of $\nabla y^{(j)}$ is supported on $K$. The Young measure gives the limiting distribution ($j \rightarrow \infty$) of the values of $\nabla y^{(j)}$ in a small neighbourhood of each point $x$. This theory uses sophisticated real analysis, and provides an elegant explanation of why fine microstructures are formed.

\begin{figure}[!ht]
\begin{minipage}{0.4\linewidth}
\centering
(a) \par\smallskip
\includegraphics[width=\linewidth]{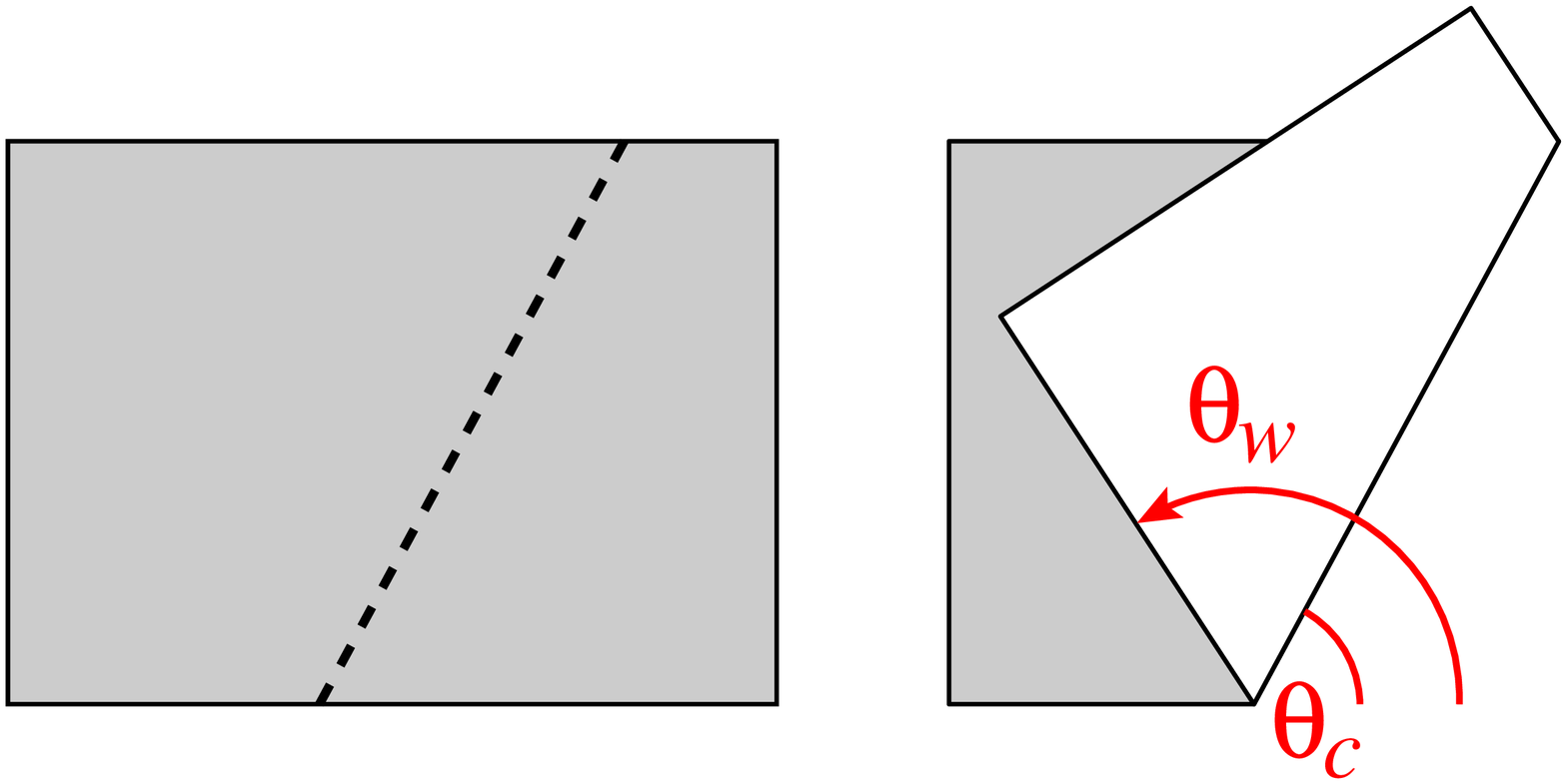}
\end{minipage}
\hspace{0.1\linewidth}
\begin{minipage}{0.4\linewidth}
(b) \par\smallskip
\includegraphics[width=\linewidth]{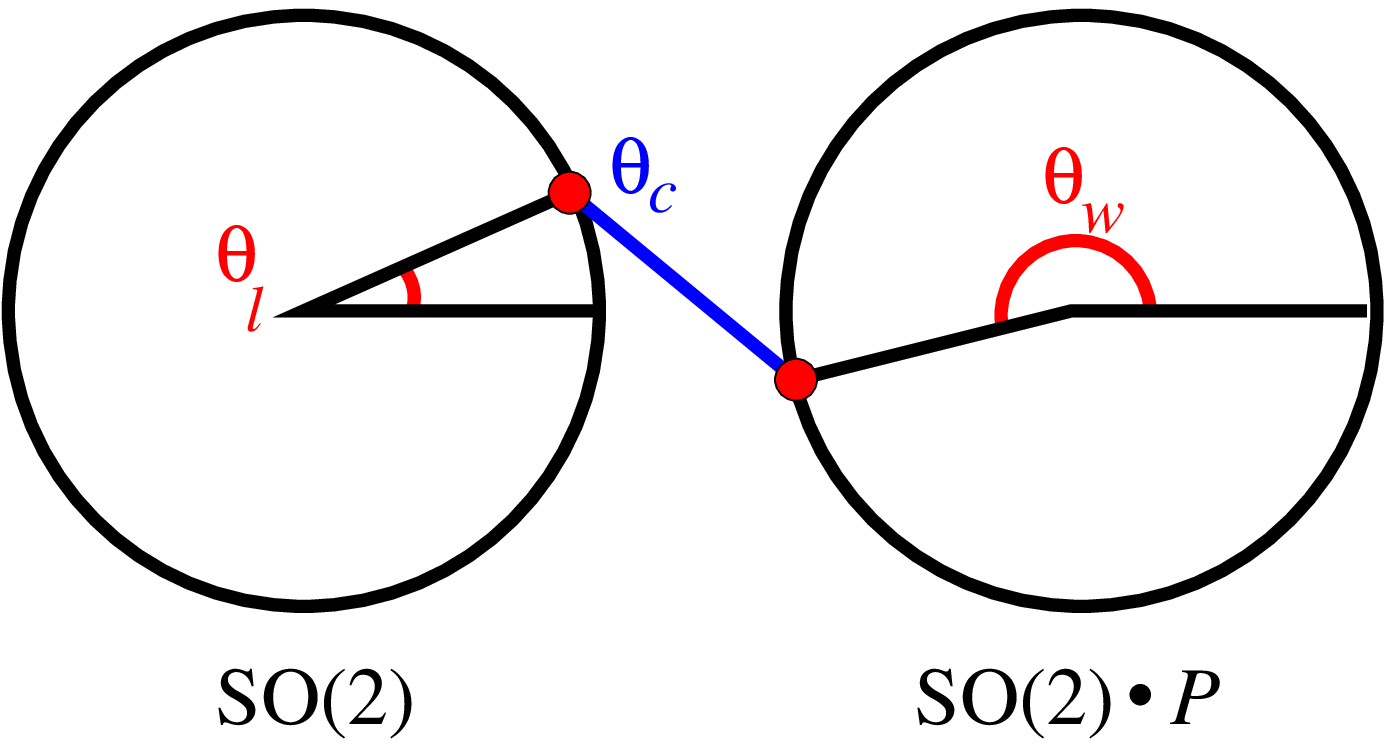}
\end{minipage}
\caption{(a) Interface between two variants of two-dimensional folded paper is a crease. (b) Order parameter space composed of two variants with compatible interfaces. See~\cite[Exercise 11.7]{Sethna06}.
\label{fig:PaperFolding}}
\end{figure}

Another instance of interesting microstructure formation, which can sometimes be regarded as a kind of toy model of the martensitic microstructure, is the folding of a piece of paper in the plane~\cite{james} (see Problems 11.7 and 11.8 of~\cite{Sethna06}). Let us consider a sheet of paper with a different color, e.g. white and grey, on each of its sides. The state of the paper, as in the martensite case, is completely characterized by the position vector field in target space $y = y(x)$, where $x$ gives the position in reference space (which we assume to have the grey face up), which in this case is in a bounded region $\Omega$ of $\mathbb{R}^2$. Two-dimensional proper rotations by an angle $\theta_l$ do not alter the paper free energy, so that the system is invariant under the action of $\text{SO}(2)$. The free energy is also invariant under a flip of the paper over the horizontal direction, achieved via an application of $P = \text{diag}(1,-1)$ in $x$, followed by a proper rotation by $\theta_\omega$, i.e. the system is invariant under the action of $\text{SO}(2) \cdot P$. Now the interface between two variants is a straight line crease with angle $\theta_c$ (Fig.~\ref{fig:PaperFolding}a). The order parameter space can be represented by two unit circles ($\mathcal{S}_1$) representing the actions of $\text{SO}(2)$ and $\text{SO}(2) \cdot P$, with compatible behaviour along the interfaces (Fig.~\ref{fig:PaperFolding}b). The two low-energy configurations, or variants, have one of the faces up, and are given by the two-dimensional identity matrix $\mathbb{I}$ and $P$. The elastic free energy density can be explicitly written as:
\begin{eqnarray}
f_{\text{p}} = \alpha \sum_{j, k} \left[ \frac{\partial y_j}{\partial x_i} \frac{\partial y_k}{\partial x_i} - \delta_{jk} \right]^2,
\end{eqnarray}
where $\alpha>0$ is a large constant, so that $f_{\text{p}}$ grows quickly when the paper is stretched or sheared. Using geometrical arguments, it is possible to show that the compatibility condition:
\begin{eqnarray}
\left( \partial_j y_i^{l} - \partial_j y_i^\omega \right) c_j = 0,
\end{eqnarray}
has to be satisfied for an interface along an axis $\hat{\bm{c}}$ separating the two variants. One can show that the paper folding example forms a microstructure that is similar to the martensite for suitable boundary conditions.

\section{Low-energy structures of four-dimensional smectic-A liquid crystals}

Dupin cyclides in $\RR^3$ can be realized as equal-time slices of certain null 3D hypersurfaces in Minkowski space $\RR^{1,3}$. In the case of a toroidal Dupin surface, $S$, such a hypersurface has as focal sets the two curves $\Sigma=\{x^2+y^2=R^2,z=\phi=0\}$ and $\bar{\Sigma}=\{x=y=0,\phi^2-z^2=R^2\}$, which sit on orthogonal planes inside $\RR^{1,3}$ (with timelike coordinate $\phi$ and $c=1$). $S$ is then given by the union of all light rays connecting $\Sigma$ to $\bar{\Sigma}$~\cite{alexander10}.

All elliptic-hyperbolic Dupin cyclides can be obtained by means of a Poincar\'e transformation acting on $S$. In fact, let us consider a Lorentz boost in the $x$-direction with ``velocity'' parameter $\beta$:
\begin{align*}
\Phi =& \gamma\, (\phi - \beta x), \\
X =& \gamma\, (x- \beta \phi), \\
Y =& y, \\
Z =& z,
\end{align*}
where $\gamma=(1-\beta^2)^{-1/2}$. It is possible to show that the transformed surface, $S'$, has equal-time surfaces corresponding to Dupin cyclides with focal curves given by the ellipse/hyperbola pair:
$$\frac{X^2}{A^2}+\frac{Y^2}{C^2-A^2}=1,
\quad\quad\quad
\frac{X^2}{C^2}-\frac{Z^2}{C^2-A^2}=1,$$
where $A=\gamma R$ and $C=|\beta|\gamma R$. The eccentricity of (both of) these conics is $e=|\beta|$. 

To put this ellipse/hyperbola pair in a generic position after applying this boost, we need 3 rotations and 4 translations (so that both the spatial origin and the $\phi$ origin are changed). In this way we need 8 generators out of the 10 generators of the Poincar\'e group. 

We can also start with the ``canonical'' toroidal Dupin cyclide with $R=1$ and use a global dilation to get an arbitrary value for $R$. The total number of parameters that are necessary to construct a generic elliptic/hyperbolic focal domain out of the ``canonical'' toroidal case is therefore 9. In this case the symmetry group should be taken as the ``Poincar\'e + dilatations'' group, which is also known as the Weyl-Poincar\'e ($\mathcal{WP}$) group. Another way of counting the number of generators is to calculate the continuous subgroup of the Weyl group which preserves both $\Sigma$ and $\bar{\Sigma}$. This is given by the Abelian subgroup $H$ of $\mathcal{WP}$ composed by rotations around $z$ and boosts in the $z$-direction. In this way, the ``space of elliptic-hyperbolic Dupin cyclides'' is intimately related to the homogeneous space given by the quotient $\mathcal{WP}/H$, which has dimension $9$.

\section{Domain decomposition and Clustering}

We developed a clustering algorithm to decompose the two-dimensional plane corresponding to a planar section of a simulated smectic liquid crystal into a set of clusters characterized by concentric circles with distinct centers. Additional details on our simulations, including dynamical behavior and visualization tools, can be found in our previous publication~\cite{liarte15}.

We start with clusters of linear size $l=32a$, where $a$ is the lattice spacing of a square grid of edge size $L=256a$. For each cluster $\alpha$, we find the four-tuple $X_\alpha = (\phi_{0,\alpha},x_{0,\alpha},y_{0,\alpha},\sigma_{0,\alpha})$ that minimizes the cost function:
\begin{eqnarray}
C_{\alpha} = \sum_{i =1}^{M_{\alpha}} c \, ( r_{i,\alpha} , X_\alpha),
\end{eqnarray}
where the triples $r_{i,\alpha} = (\phi_{i,\alpha}, x_{i,\alpha}, y_{i,\alpha})$ are coordinates (in (2+1)D space) of the points belonging to cluster $\alpha$, $1 \leq i \leq M_\alpha$ with $M_\alpha$ denoting the size of the cluster, and
\begin{eqnarray}
c\, ( r_i , X_\alpha)
 &=& \left\{
\phi_i - \phi_{0,\alpha}
+ \sigma_{0,\alpha}
\left[ \left(x_i - x_{0,\alpha} \right)^2 
+ \left(y_i - y_{0,\alpha} \right)^2 \right]^{1/2} \right\}^2
\nonumber \\ && \quad
+ \left( N_x - \frac{x_i - x_{0,\alpha}}{d_{i\,0,\alpha}}\right)^2
+ \left( N_y - \frac{y_i - y_{0,\alpha}}{d_{i\,0,\alpha}}\right)^2,
\label{eq:Cost}
\end{eqnarray}
where $d_{i\,0,\alpha}=[(x_i-x_0^{(\alpha)})^2+(y_i-y_0^{(\alpha)})^2]^{1/2}$ is the two-dimensional Euclidean distance from point $i$ to the center $(x_0,y_0)$, and $\sigma_0^{(\alpha)} =\pm 1$ characterizes the nappe of the cone, being positive (negative) for the set of points lying below (above) the apex. By minimizing the first term in the right-hand side of Eq. \eqref{eq:Cost}, we find the best approximation for one of the local energy structures given by Eq. [5] of the main text. The second and third terms make the analysis sensitive to gradient changes. The numerical implementation is performed using a k-means least-squares optimization algorithm.

After we find the best four-tuples $X_\alpha$ for each cluster $\alpha$, we reassign each three-dimensional vector $r_i= (\phi_{i}, x_{i}, y_{i})$ to the cluster $\alpha$ that yields the smallest value of $c\,(r_i, X_\alpha)$. We then iterate this process (finding best parameters and redefining clusters) several times (one hundred times for the plot displayed in Fig. 4 in the main paper). In each iteration we discard clusters of size smaller than four, in order to ensure numerical convergence.

A pair of clusters $\alpha$ and $\beta$ that are very similar, in the sense that they yield similar four-tuples $X_\alpha$ and $X_\beta$, are combined into a single cluster $\alpha \cup \beta$, which is characterized by new parameters $X_{\alpha \cup \beta}$. More precisely, we implement these ideas by defining a matrix $\Delta C = (( \Delta C_{\alpha \, \beta}))$, with
\begin{eqnarray}
\Delta C_{\alpha \, \beta} = \frac{C_{\alpha\, \cup\, \beta } - C_\alpha - C_\beta}{\min (M_\alpha, M_\beta)},
\end{eqnarray}
where $\alpha$ and $\beta$ run over all cluster indices. We merge the pair of clusters $\alpha^*$ and $\beta^*$ with the smallest $\Delta C_{\alpha \, \beta}$, for $\alpha < \beta$, if $\Delta C_{\alpha^* \,\beta^*}$ is smaller than a certain threshold. We then rebuild $\Delta C$ and repeat this process several times (we did twenty such iterations in the plot of Fig. 4 of the main paper).

Finally, each pixel in Fig. 4 is colored according to its cluster, defined by the set of four-tuples $X_\alpha$. We use $\{ \cos^2(2\pi x_0 / L),\sin^2(2\pi y_0 / L),\cos^2(2\pi \phi_0 / (2L) ) \}$ in RGB scale.

%

\end{document}